\documentstyle[epsfig,epsf,preprint,amsbsy,amssymb,aps]{revtex}

\def\one{{\mathchoice {\rm 1\mskip-4mu l} {\rm 1\mskip-4mu l} {\rm
1\mskip-4.5mu l} {\rm 1\mskip-5mu l}}}
\newcommand{\ket}[1]{|{#1}\rangle}
\newcommand{\bra}[1]{\langle{#1}|}

\renewcommand{\tensor}{\otimes}

\newcommand{\lowertr}{\mbox{lt}}
\newcommand{\diag}{\mbox{dg}}

\begin{document}

\draft
\title{
Quantum Algorithms for Fermionic Simulations}
\author{G. Ortiz, J.E. Gubernatis, E. Knill, and R. Laflamme}
\address{
Los Alamos National Laboratory, Los Alamos, NM 87545}

\date{\today}
\maketitle

\begin{abstract}
We investigate the simulation of fermionic systems on a quantum
computer. We show in detail how quantum computers avoid the dynamical
sign problem present in classical simulations of these systems,
therefore reducing a problem believed to be of exponential complexity
into one of polynomial complexity. The key to our demonstration is the
spin-particle connection (or generalized Jordan-Wigner transformation)
that allows exact algebraic invertible mappings of operators with
different statistical properties.  We give an explicit implementation
of a simple problem using a quantum computer based on standard qubits.
\end{abstract}

\pacs{Pacs Numbers: 3.67.Lx, 5.30.-d,}
\columnseprule 0pt

\narrowtext

\section{Introduction}
\label{section1}

Because of recent exciting algorithms, like the factoring algorithm of
Shor \cite{shor:qc1995a} and the search algorithm of Grover
\cite{grover:qc1997a}, that solve difficult problems on a quantum
computer using algorithms that would be impractical on a classical
computer, it is easy to overlook that the original proposals for
quantum computers were for the purpose of solving quantum physics problems
\cite{feynman:qc1982a}. People like Feynman \cite{feynman:qc1982a}
focused on the extent to which such a computer could imitate a specific
physical process, suggesting in part that quantum problems were
inherently too complex for a classical computer \cite{feynman:qc1982a}.

The obvious difficulty with deterministically solving a quantum
many-body problem (of fermions or bosons) on a classical computer is
the exponentially large basis set needed (i.e., the dimension of its
Hilbert space grows exponentially with the number of degrees of
freedom). Exact diagonalization approaches (e.g., the Lanczos method)
suffer from this exponential ``catastrophe''. Viewed the other way
around, this basis set scaling is what restricts today's classical
computer to simulating only small quantum computers.  This point seems
indisputable, but should not be taken as proof that quantum systems
cannot be simulated on a classical computer. By the same token,
the
recent claim \cite{abrams:qc1997a} that quantum computers can simulate
all quantum systems efficiently lacks explicit and detailed algorithms
for specific problems, and lacks a generic model of quantum computation
including the unitary maps (quantum gates) that can be physically
implementable.
Even if a quantum computer existed, some interesting quantum problems,
such as finding the ground state of a general quantum Hamiltonian, do not yet
have efficient quantum algorithms. Finding such a quantity for small
systems is relatively routine on a classical computer.

Feynman in fact analyzed two alternatives for simulating physics with
computers \cite{feynman:qc1982a}. One uses a probabilistic classical
computer that would produce from the same input as given to a physical
system the same distribution of outputs as observed for the physical
system. The other uses a computer constructed of distinctively quantum
mechanical elements that obey the laws of quantum mechanics. This
latter proposal is the quantum computer.

To the question, ``Can quantum systems be probabilistically
simulated by a classical computer?'', Feynman's answer was
unequivocally ``No.''~\footnote{There is as yet no mathematical
proof that this is the correct answer.}  This answer is surprising
for even at that time some quantum systems were being very
successfully simulated probabilistically on classical computers,
mainly by quantum Monte Carlo (QMC) methods \cite{kalos:qc1974a}.
To the question, ``Can quantum systems be simulated with a quantum
computer?'', his answer was a qualified ``Yes.'' He believed
almost certainly that this could be done for a system of bosons
but was unsure that it could be done for a system of fermions. In
this paper we present a design for a universal quantum computer
that will simulate a system of fermions. Before doing so, we first
discuss some problems that can be solved by a probabilistic
simulation of a quantum system on a classical computer and others
that cannot.

Probabilistic simulations of quantum systems on a classical computer
are mainly performed with the use of the Monte Carlo method. These
statistical approaches were introduced to overcome the difficulty of
exponentially growing phase spaces by numerically evaluating the
accompanying many-dimensional integrals by sampling from a function
assumed to be non-negative. On a classical computer one can
probabilistically simulate a quantum system like liquid He$^4$
\cite{ceperley:qc1995a} and produce results that accurately compare
with experiment. The situation, however, is far from satisfactory. An
unsatisfactory state of affairs results from the frequent breakdown of
the non-negativity assumption and is called ``the sign problem.'' The
sign problem is manifested by the seemingly exponentially hard task of
estimating the expectation value of an observable with a given error.
Interestingly, Feynman's negativism about quantum systems being
probabilistically simulated by classical computers was a claim that
negative probabilities were unavoidable because of the ``hidden
variable'' problem and the possible violation of Bell
inequalities. The extent to which the sign problem is a hidden
variable problem is unclear. On the other hand, QMC methods do not
faithfully adhere to Feynman's idea of a probabilistic computer.  Two
important differences are that most QMC simulations are non-local and
performed in imaginary time. Feynman discussed real-time simulations
on a local computer. Implications of these differences have been noted
by Ceperley \cite{david} who suggests Feynman really argues just
against simulating quantum dynamics on a local classical computer. In
any case, probabilistic simulations on a classical computer clearly do
not qualify as a {\it universal} computational scheme for general
quantum many-body problems. The limiting factors, for whatever
reasons, are negative or complex-valued probabilities whether the
simulations are done in real or imaginary time.

To place the sign problem in a better perspective, we will start with a
real-time analysis of a collection of interacting quantum particles.
Quantum mechanics tells us that these particles either obey Bosonic
statistics, whereby the wave function is symmetric with respect to the
exchange of the states of any two particles, or obey Fermionic
statistics, whereby the wave function is antisymmetric (changes sign)
with respect to the exchange of any two particles \cite{anyons}. Examples
of bosons are photons and gluons; examples of fermions are electrons,
protons, neutrons, and quarks. Often these two quantum statistics
conveniently and efficiently map onto a third, quantum spin statistics
\cite{ours}. Still in other cases, when particle exchange is unlikely,
particle statistics is simply ignored.

For a given initial quantum state $|\Psi(0)\rangle$, a quantum computer
solves the time-dependent Schr\"odinger equation
\begin{equation}
 i\hbar \frac{\partial|\Psi\rangle}{\partial t} = H|\Psi\rangle
\end{equation}
by incrementally propagating the initial state via
\begin{equation}
  |\Psi(t)\rangle
   = \underbrace{e^{-i\Delta t H/\hbar} e^{-i\Delta t H/\hbar}\cdots
     e^{-i\Delta t H/\hbar}}_{M \ {\rm factors}} |\Psi(0)\rangle \ .
\end{equation}
($t= M \Delta t$ and the Hamiltonian $H$ is assumed time independent).
It should be reasonably apparent that if the Monte Carlo method is
applied to the evaluation of the right-hand side of this equation, it
is faced with sampling from oscillatory integrands that are not always
positive and have unknown nodal surfaces. Further, as time $t$
increases, the integrand fluctuates with increasing rapidity. While
clever stationary-phase forms of the QMC method have been developed,
acceptable solutions are possible only for relatively short
times. This form of the sign problem is called the {\it dynamical sign
problem}, and we are unaware of any efficient \cite{Note1} real-time
QMC simulations for bosonic, fermionic, or quantum spin systems.

Years ago \cite{kalos:qc1974a}, before quantum computers were proposed,
it was realized that by transforming Schr\"odinger's equation to
imaginary-time $\tau$ via $t\rightarrow -i\hbar\tau$ the problem with
the rapid fluctuations was eliminated. With this transformation, called
Wick's rotation, one solves the diffusion-like equation
\begin{equation}
 \frac{\partial|\Psi\rangle}{\partial \tau} = -H|\Psi\rangle
\end{equation}
by incrementally propagating the initial state via
\begin{equation}
 |\Psi(\tau)\rangle
 = \underbrace{e^{-\Delta\tau H}e^{-\Delta\tau H}\cdots e^{-\Delta\tau
 H}}_{M \ {\rm factors}} |\Psi(0)\rangle \ .
\end{equation}
($\tau= M \Delta \tau$ and the Hamiltonian $H$ is assumed time
independent.) This transformation permits QMC simulations of
time-reversal invariant interacting boson systems to a high degree of
accuracy. For systems of interacting quantum spins and fermions (or
bosons with complex hermitian Hamiltonians \cite{ortiz:qc1993a}), the
transformation does not solve the sign problem.  For quantum spin
systems, the difficulty is finding a basis in which all matrix elements
of the positive-definite operator $\exp({-\Delta\tau H})$ are positive.
Most notably this difficulty occurs for frustrated quantum spins. For
fermion systems, the problem is the Monte Carlo process causing state
exchanges that because of the anti-symmetrization requirement just
happen to produce samples which are as frequently positive as negative.
For the sign problem found in both types of systems, the statistical
error of the measured observables grows exponentially fast with
increasing  system size. Another difficulty with the imaginary-time
approach is analytically continuing the results back to real-time if
real-time, i.e., truly dynamical, information is needed \cite{janez}.
This continuation is an ill-posed problem whose solution places
extraordinary demands on the simulation \cite{jarrell:qc1996a}.

In this paper, we will focus on the dynamical sign problem for a system
of fermions, seemingly the most challenging case. Eventually we will
give a detailed implementation of a simulation of the dynamical
properties of a collection of interacting fermions on a quantum
computer. The implementation avoids the sign problem. First, in Section
\ref{section2} we will discuss more fully the mathematical origin of
the dynamical sign problem in classical computation and show why a
quantum algorithm overcomes the problem. In Section \ref{section3} we
will give the elements required for Deutsch's quantum network model of
a quantum computer \cite{deutsch:qc1985a}. The quantum gate in this
model conveniently allows the propagation of systems of local two state
objects, e.g., a localized quantum spin-$\frac{1}{2}$ particle called
qubit.
We also propose a universal set of quantum gates (unitary operators)
that allows generic propagation of systems of fermions (the fabled
``Grassmann chip'' \cite{chodos:qc1983}). The resulting fermion algebra
has been the main technical tool for studying the classical Ising model
in two spatial dimensions \cite{onsager}, a prototype lattice model
that had an enormous impact on our understanding of phase transitions.
Next, In Section~IV,  we show how this propagation can be effected by
the quantum spin gate. We will demonstrate the polynomial scaling of
the construction of the initial state, its subsequent time propagation,
and the measurement of some observable. Here we will also demonstrate
the control of the error in the results. In Section \ref{section4}, we
apply our model of dynamical fermion computation to a toy problem to
illustrate our procedures in more detail. The exact solution of this
toy problem will be given in the Appendix; however, if one were to
solve this problem in the obvious Hilbert space, its solution would
require a quantum computer for a sufficiently large system. Finally, in
Section \ref{section5}, we summarize and make some remarks about future
research directions.

Our universal fermion gate and its mapping to the standard universal
quantum gate is very similar to the one recently discussed by Bravyi
and Kitaev \cite{bravyi:qc2000a} who actually propose that a quantum
computer built from fermions might be more efficient than one built
from distinguishable two state systems.

\section{Dynamical Sign Problem}
\label{section2}

In order to understand the mathematical origin of the dynamical sign
problem we use the Feynman path integral formulation
\cite{feynman:qc1965a} for continuum systems in the first quantization
representation. In this formalism one maps a quantum problem in $D$
dimensions into a classical one in $D+1$ dimensions and then simulates
that problem probabilistically on a classical computer. The algorithm
is efficient except for the repetition needed to obtain sufficiently
good statistics.  The ``distinguishable particle'' quantum mechanical
propagator of a system represented by the Hamiltonian $H=\frac{1}{2}
\sum_{i=1}^{N_e} p_i^2 + V({\cal R})$ is expressed as
\cite{Note0}
\begin{equation}
G({\cal R} \rightarrow {\cal R}^\prime; t) = \langle {\cal R}', t | e^{-i
H t} | {\cal R}, 0 \rangle = \int_{{\cal R}(0)={\cal
R}}^{{\cal R}(t)={\cal R}'} {\cal D} [{\cal R}(t)] \
e^{i S \left[{\cal R}(t) \right] } \ ,
\label{prop}
\end{equation}
where the measure ${\cal D} [{\cal R}(t)] =  \lim_{M \rightarrow
\infty} (2 \pi i t/M)^{-M D/2} \ d{\cal R}_1 \cdots d{\cal R}_{M-1}$,
and the action
\begin{equation}
S \left[{\cal R}(t) \right] = \int_0^t d \tau \left \{ \frac{1}{2}
\left ( \frac{d {\cal R}(\tau)}{d \tau} \right )^2 - V({\cal R}(\tau))
\right\} \ .
\end{equation}
Bosonic or fermionic statistics are introduced by applying the
corresponding symmetrization operator to the propagator, Eq.
\ref{prop}. However, because the dynamical sign problem occurs for any
particle statistics, we will ignore particle statistics for the sake
of simplicity.

The description of the properties of different physical systems in
terms of correlations of physical observables is the natural way to
compare with available experimental information. In this regard,
linear response theory provides a way to compute the response of a
system to a weak external dynamical perturbation \cite{negele:qc1988a}.
This linear response is always expressed in terms of a time correlation
function of the dynamical variables that couple to the perturbation.
For example, if we were to apply an external time-dependent magnetic
field and we wanted to calculate the average induced magnetization, we
would have to compute a time-dependent magnetization-magnetization
correlation function. The two-time correlation function between
arbitrary local dynamical variables $A$ and $B$ is
\begin{equation}
C_{AB}(t) = \langle A(t) B(0) \rangle = \langle e^{i H t} A e^{-i H t} B
\rangle \ ,
\end{equation}
if the Hamiltonian is time independent. Expressed in integral form
\begin{equation}
C_{AB}(t) = \frac{\int d{\cal R}d{\cal R}'d{\cal R}'' \ \rho({\cal
R},{\cal R}'') G({\cal R}' \rightarrow {\cal R}''; -t) A({\cal R}')
G({\cal R} \rightarrow {\cal R}'; t) B({\cal R})}{\int
d{\cal R}d{\cal R}'d{\cal R}'' \ \rho({\cal
R},{\cal R}'') G({\cal R}' \rightarrow {\cal R}''; -t)
G({\cal R} \rightarrow {\cal R}'; t)} \ ,
\label{cab}
\end{equation}
where the ``distinguishable particle'' density matrix $\rho$ specifies
the way the system was initially prepared. The next step consists of
writing Eq. \ref{cab} in path-integral form using Eq. \ref{prop}
\begin{equation}
C_{AB}(t) = \frac{\int d{\cal R}_1 \cdots d{\cal R}_{2M} \ A({\cal
R}_{M+1})  B({\cal R}_{1}) \ P({\cal R}_1, \cdots, {\cal R}_{2M}) \ e^{i
\Phi({\cal R}_1, \cdots, {\cal R}_{2M})}}{\int d{\cal R}_1 \cdots
d{\cal R}_{2M} \ P({\cal R}_1, \cdots, {\cal R}_{2M}) \ e^{i
\Phi({\cal R}_1, \cdots, {\cal R}_{2M})}} \ ,
\label{cab1}
\end{equation}
where $P$ and $\Phi$ are real-valued functions. Generically,  a
stochastic estimate of $C_{AB}(t)$ is
\begin{equation}
C_{AB}(t) = \frac{\sum_{\{{\cal R}_i\}} A({\cal R}_{M+1})  B({\cal
R}_{1}) \ e^{i \Phi(\{{\cal R}_i\})}}{\sum_{\{{\cal R}_i\}} \ e^{i
\Phi(\{{\cal R}_i\})}} = \frac{\langle A({\cal R}_{M+1})  B({\cal R}_{1})
e^{i \Phi(\{{\cal R}_i\})} \rangle_P}{\langle e^{i \Phi(\{{\cal
R}_i\})} \rangle_P} ,
\label{cab2}
\end{equation}
where the configurations $\{ {\cal R}_i \}$ are sampled from the
probability distribution $P$ (positive semidefinite measure). One
immediately sees that the origin of the dynamical sign problem is the
oscillatory phase factor $e^{i \Phi}$ that leads to large phase
fluctuations at long times. Manifestly, $|\langle e^{i \Phi(\{{\cal
R}_i\})} \rangle_P| \rightarrow 0$ in an exponential fashion as $t$
gets larger. Therefore, the total statistical error for the evaluation
of $C_{AB}(t)$ grows exponentially with time because of large
cancellations both in the numerator and denominator. The so-called
``fermion sign problem'' is a particular case of this problem when
$e^{i \Phi}=\pm 1$ and time is imaginary \cite{vonderlinden:qc1992a}.

Will a quantum computer solve this problem? At first glance it appears
that it will. A quantum computer is a physical system, whether a system
of fermions or not, and physical systems have no dynamical or fermion
sign problems. Furthermore it has been argued that there are means for
mapping most physical systems to a quantum computer in such a way that
the quantum computer's controlled evolution mimics that of the physical
system~\cite{feynman:qc1982a,lloyd:qc1996a}. A closer look, however,
makes the situation less clear. A quantum computer is a computer, and
as such it suffers from limited accuracy. More importantly this type of
computer predicts results stochastically, meaning each measurement is
one member of the ensemble of measurements possible from a distribution
specified by the modulus squared of the wave function for the
Hamiltonian $H$ modeled by the quantum computer. For a fixed physical
time $t>0$, how accurate is an individual measurement, how accurate is
the expectation value of these measurements, and how controlled is
their estimated variance? Is the level of accuracy and control
achievable polynomially with complexity and $t$?

There is an area where a problem similar to the sign problem
has been recognized and resolved by quantum computation. Recently
it was shown that quantum computation is polynomially equivalent
to classical probabilistic computation with an oracle for
estimating the value of simple sums of rational numbers called {\em
quadratically signed weight enumerators\/} (QWGTs)
\cite{knill:qc1999c}. In other words, if these sums could be
evaluated, one could use them to generate the quantum statistics
needed to simulate the desired quantum system. More specifically,
what was demonstrated was the obtainability of expectation value
of operators in quantum computation by evaluating sums of the form
\begin{eqnarray}
S(A,B,x,y) &=& \sum_{b:Ab=0}(-1)^{b^TBb}x^{|b|}y^{n-|b|},
\end{eqnarray}
where $A$ and $B$ are $0$-$1$-matrices with $B$ of dimension
$n \times n$ and $A$ of dimension $m \times n$. The variable $b$
in the summand ranges over $0$-$1$-column vectors of dimension
$n$, $b^T$ denotes the transpose of $b$, $|b|$ is the {\em
weight\/} of $b$ (the number of ones in the vector $b$), and all
calculations involving $A$, $B$ and $b$ are done modulo $2$. The
absolute value of $S(A,B,x,y)$ is bounded by $(|x|+|y|)^n$.
Quantum computation corresponds to the problem of determining the
sign of $S(A,\lowertr(A),k,l)$ with the restrictions of having
$\diag(A)=I$, $k$ and $l$ being positive integers, and
$|S(A,\lowertr(A),k,l)| \geq (k^2+l^2)^{n/2}/2$. $\diag(A)$ is a
diagonal matrix formed from the diagonal elements of $A$ and
$\lowertr(A)$ is a lower triagonal matrix formed from the lower
triangular elements of $A$.  Details of this quantum algorithm can
be found in \cite{knill:qc1999c}.

The main point is that these sums are similar to the numerator of
Eq.~\ref{cab2}, and attempts to estimate them by random sampling
result in exponentially bad signal to noise ratios. In the case of
QWGTs, quantum computers can estimate the sum exponentially better
than classical computers, but the estimate is not exact. The
situation for the dynamical sign problem is similar: Quantum
computers cannot obtain exact values for the desired correlation
functions, but can obtain estimates sufficiently exact to avoid
the sign problem suffered by the known classical algorithms and to
yield usable information about the physical models simulated.

In this paper we will show explicitly how the sign problem is avoided
in the case of simulating fermions. Below we will give a means for
translating local fermion Hamiltonians into the Hamiltonians available
in the standard model of quantum computation. In contrast to quantum
simulations on a classical computer this translation prevents
uncontrolled exchange processes that are the dominant source of the
fermion sign problem. With respect to the dynamical sign problem, we
then argue by using standard error correction analysis developed for
the standard model of quantum computing that these gates will enable
sufficiently accurate measurements of correlation functions so the
accuracy of the average of these measurements will be dominated by the
statistical error. The statistical error is problem dependent but
polynomially bounded, so that the difficulty associated with
phase-weighted averages is eliminated.

\section{Models of Quantum Computation}
\label{section3}

The {\em quantum control} model of quantum computation assumes the
existence of physical systems that can be controlled by modulating
the parameters of the system's Hamiltonian $H_P$. The control
possibilities are abstracted and used to implement specific
\emph{quantum gates} that represent the unitary evolution of the
physical system over a time step obtained by specific modulations
of the Hamiltonian. In most treatments, the physical systems,
together with the gates, are then taken as the abstract model of
quantum computation. The quantum control and quantum gate
viewpoints are effectively equivalent, but to tie the
computational model to the physics simulation problem more
closely, we choose to describe quantum computation from the point
of view of quantum control; that is, we will assume an $H_P$. In
this context we begin by giving the standard model of quantum
computation and then defining an alternative model based on
fermions.

Defining a model of quantum computation consists of giving an
algebra of operators, a set of controllable Hamiltonians
(Hermitian operators in the algebra), a set of measurable
observables, and an initial state of the physical system. In the
simplest case, the observables are measured by the method of
projective measurements, and the initial state of the physical
system is an expectation value of the algebra's operators.

\subsection{Standard Model of Quantum Computation}

The standard model of quantum computation (Deutsch's quantum
network representation) is based on an assembly of two state
systems called \emph{qubits}, controlled by one- and two-qubit
Hamiltonians, and on a measurement process determined by one-qubit
observables.

\noindent{\bf Operator algebra:} It is convenient to define the
standard model through the algebra of operators acting on the
qubits. This algebra is generated by the unit and Pauli matrices
$\sigma_x$, $\sigma_y$ and $\sigma_z$ for each qubit $j$,
\begin{equation}
\one =
\pmatrix{1&0 \cr 0&1 \cr} \ ; \
\sigma_x= \pmatrix{ 0&1 \cr 1&0 \cr} \ ; \
\sigma_y=\pmatrix{ 0&-i \cr i&0 \cr} \ ; \
\sigma_z=\pmatrix{ 1&0 \cr 0&-1 \cr} \ .
\end{equation}
These matrices represent quantum operators with mixed commutation
relations and span the space of complex-valued $2\times 2$
matrices. For qubits $j \neq k$, the $\sigma$'s commute, and for
qubits $j = k$, they satisfy the relation
$\sigma_\mu\sigma_\nu+\sigma_\nu\sigma_\mu = 2\delta_{\mu\nu}
\one$, ($\mu,\nu = x,y,z$).  For a \emph{quantum register} with
$n$ qubits, one may take the operator $\sigma^j_\mu$ defined in
terms of a Kronecker product
\[
\sigma^j_\mu = \one \otimes \one \otimes \cdots \otimes
\underbrace{\sigma_\mu}_{j^{th} \ {\rm factor}} \otimes \cdots \otimes \one 
\]
of matrices acting on $n$ two-dimensional linear spaces.  Thus
$\sigma^j_\mu$ admits a matrix representation of dimension $2^n \times
2^n$.

\noindent{\bf Control Hamiltonians:}
Control of qubits is achieved by applying Hamiltonians that act on
either one or two qubits. A theorem
\cite{barenco:qc1995a,divincenzo:qc1995a} in quantum information
processing says that a generic operation on a single qubit and any
interaction between two qubits is sufficient for building any unitary
operation. We take
\[
H_P(t)=\sum_j [\alpha_{x_j}(t) \ \sigma_x^j  + \alpha_{y_j}(t) \ \sigma_y^j]
+\sum_{i,j} \alpha_{ij}(t) \ \sigma_z^i \sigma_z^j,
\]
where the $\alpha_{\mu}(t)$ are controllable. Ideally, no
constraints on the control functions are assumed. However, it is
often simpler to design the required control by assuming that only
one of the $\alpha_{\mu}(t)$ is non-zero at any time. A quantum
algorithm for this model of quantum computation consists of
prescribing the control functions~\cite{Note2}. A convenient
measure of the complexity of such an algorithm is the integral
$\int_{0}^{t} dt' \sqrt{\sum_\mu \alpha^2_{\mu}(t')}$ (the
\emph{action} of the algorithm). The quantum gates are simply
specific unitary evolutions that may be implemented in terms of
$H_P$. A convenient universal set of gates is given by operators
of the form $\exp(i\sigma^i_\mu\pi/4)$ and
$\exp(i\sigma^i_z\sigma^j_z\pi/8)$.
In the quantum network
representation of the standard model, an algorithm is a specific
sequence of these operators applied to the initial state of the
qubits.

\noindent{\bf Initial state:} The initial state of the qubits is
assumed to be an $n$ term Kronecker product of the state
$|0\rangle\equiv {1 \choose 0}$ which is an eigenstate of
$\sigma_z$with eigenvalue $1$. The state is completely determined
by the expectation values $\langle 0| \sigma^i_{\mu_i} |0\rangle$,
which are $1$ if the $\sigma^i_{\mu_i}$ are all $\sigma^i_z$ or the
identity, and are $0$ otherwise. Physically, the initial state has
all ``spins'' up.

\noindent{\bf Measurement:} The final feature of the model of
computation is the specific means for extracting information after
a sequence of operations has been applied to the initial state. In
the standard model, it is always possible to apply a projective
(von Neumann) measurement \cite{peres:qc1998a} using the
observables $\sigma^i_z$. With this capability, it is unnecessary
to give an initial state explicitly, as the desired state can be
prepared by using measurement and operations. To learn the
expectation of an observable at the end of an algorithm, one
repeats the algorithm and measurement procedure many times and
averages over the measurements until the desired accuracy is
achieved.

For a description of the standard model of quantum computation in terms
of quantum Turing machines, see~\cite{bernstein:qc1997a}. Quantum
networks are discussed in~\cite{barenco:qc1995a}. Introductory
descriptions of the standard model may be found
in~\cite{cleve:qc1999b,aharonov:qc1998b}.

\subsection{Fermion Model of Quantum Computation: Grassmann
chip}

Somewhat analogously, we now describe a standard model of fermion
computation. For simplicity we only consider spinless fermions,
i.e., fermions without internal spin degrees of freedom, although
we could have considered more general fermionic algebras with
internal degrees of freedom \cite{ours}. Physically, a system of
spinless fermion might be a system of spin-$\frac{1}{2}$ electrons
in a magnetic field sufficiently strong to polarize it fully. The
basic system of this model is a state (or fermionic mode) that can
be occupied by 0 or 1 spinless fermion. We define the model for
$n$ such modes.

\noindent{\bf Operator algebra:} We define the model through the
algebra of the spinless fermion operators $a_j^{\;}$ and
$a_j^\dagger$ for each qubit $j$ ($j=1,\cdots,n$), i.e., through
the algebra of $2n$ elements satisfying canonical anticommutation
relations
\begin{eqnarray}
\{a_i^{\;},a_j\} = 0 \; , \;
\{a_i^{\;},a_j^{\dagger}\} = \delta_{ij} \nonumber \ ,
\end{eqnarray}
where $\{A,B\}= AB+BA$ denotes the anticommutator or Jordan product.
$a_j^\dagger$ ($a_j^{\;}$) creates (annihilates) a spinless fermion in
state (mode) $j$. Each element admits a matrix representation of
dimension $2^n \times 2^n$. The fermion algebra is isomorphic (as a
$*$-algebra) to the standard model (or Pauli) algebra. The isomorphism
is established through the Jordan-Wigner mapping \cite{jordan}. 

\noindent{\bf Control Hamiltonians:}
We take
\[
H_P=\sum_j \left[\alpha_j(t) a_j^{\;} + \tilde{\alpha}_j(t)
a_j^{\dagger}\right] + \sum_{ij} \alpha_{ij}(t) \left (a_i^\dagger
a_j^{\;} + a_j^\dagger a_i^{\;}  \right) \ .
\]
This is a universal Hamiltonian, i.e., any other Hamiltonian for a
system of interacting spinless fermions can be generated by it.
Physical operators must be (Hermitian) products of bilinear
combinations of the creation and annihilation operators, i.e., products
of sums of $a_j^\dagger a_i^{\;}$ for arbitrary $i$ and $j$ modes.

\noindent{\bf Initial state:} The initial state is assumed to
be an $n$ term Kronecker product of the state $|0\rangle$
which is an eigenstate of the number operator $a_j^\dagger
a_j^{\;}$ with eigenvalue 0.  The state is completely determined by
the expectation values $\langle 0| a_j^\dagger a_j^{\;}
|0\rangle=0$ for all $j$. Physically, the initial state has all
modes unoccupied.

\noindent{\bf Measurement: } Measurements can again be performed by
using von Neumann's scheme of projective measurements. In Section
\ref{measurement}, we will discuss another scheme more appropriate for
the physical systems at hand.

In the next subsection we show how to simulate the fermion model
by using the standard spin-$\frac{1}{2}$ model. In particular it
is possible to efficiently map the fermion Hamiltonians to Pauli
operators which can be simulated using the control Hamiltonians of
the standard model. This establishes that these two models of
computation are polynomially equivalent. Here the point of view is
similar to the one used for classical models of computation: the
simulation of one model by another establishes their equivalence.

\section{Fermion Computation via the Standard Model}
\label{fermionstandard}
In the previous Section we gave the elements required for Deutsch's
quantum network model of a quantum computer \cite{deutsch:qc1985a} and
proposed a universal set of quantum gates (unitary operators) that
allows generic propagation of systems of fermions (the fabled
``Grassmann chip'' \cite{chodos:qc1983}).
Here we show how this propagation can be effected by the quantum
spin gate. We will demonstrate the polynomial scaling of the
construction of the initial state, its subsequent time propagation, and
the measurement of some observable. We will also demonstrate the
control of the error in the results.

The first step is the observation that the set of 2$n$ matrices
$\gamma_{\mu}$ (of dimension $2^n \times 2^n$) satisfying the Clifford
algebra identities
\[
\left \{ \gamma_{\mu}, \gamma_{\nu} \right \} = 2 \delta_{\mu \nu}
\]
admits a representation in terms of Pauli matrices (Brauer-Weyl
construction)
\begin{eqnarray}
\gamma_{1} = \sigma_x^1,\,  && \gamma_2 = \sigma_y^1 \nonumber \\
\gamma_{3} = \sigma_z^1 \sigma_x^2,\, && \gamma_4 = \sigma_z^1 \sigma_y^2
\nonumber \\
\gamma_{5} = \sigma_z^1 \sigma_z^2 \sigma_x^3 ,\,&& \gamma_6 = \sigma_z^1
\sigma_z^2 \sigma_y^3 \nonumber \\
\vdots  &&      \nonumber \\
\gamma_{2n-1} = [\prod_{j=1}^{n-1} \sigma_z^j ] \sigma_x^n  ,\,&&
\gamma_{2n} = [\prod_{j=1}^{n-1} \sigma_z^j ] \sigma_y^n   \ . \nonumber
\end{eqnarray}
The following mapping of fermion operators
\begin{eqnarray}
a_j^{\;} &\rightarrow& \left( \prod_{i=1}^{j-1} -\sigma^i_z \right)
\sigma^j_- = (-1)^{j-1} \ \sigma^1_z \sigma^2_z \cdots \sigma^{j-1}_z
\sigma^j_- = (-1)^{j-1} \ \frac{\gamma_{2j -1} - i \gamma_{2 j}}{2}
\nonumber \\
a^{\dagger}_j &\rightarrow& \left( \prod_{i=1}^{j-1} -\sigma^i_z
\right) \sigma^j_+ = (-1)^{j-1} \ \sigma^1_z \sigma^2_z \cdots
\sigma^{j-1}_z \sigma^j_+  = (-1)^{j-1} \ \frac{\gamma_{2j -1} + i
\gamma_{2 j}}{2} \nonumber
\end{eqnarray}
where $\sigma^j_{\pm} = \frac{\sigma^j_x \pm i \sigma^j_y}{2}$,
defines a $*$-algebra isomorphism to the algebra of operators of the
standard model. It is the so-called spin-$\frac{1}{2}$ Jordan-Wigner
transformation \cite{jordan}, and has the property that
$\hat{n}_j=a^{\dagger}_j a^{\;}_j = \sigma^j_+ \sigma^j_-$. We note
that $\hat{n}_j$ is a ``local'' particle number (or density) operator
and many types of interaction in physical systems are of the form
``density times density'' which simplifies the simulation as we will see.

It is important to emphasize that the success of our approach depends
upon the mapping of algebras (and not of Hilbert spaces). In this
regard it is relevant to mention that the transformation just presented
is a particular case of a more general set of mappings that we would
like to name generalized Jordan-Wigner transformations \cite{ours}. It
is possible to imagine a quantum computer implemented, for example,
with a three state unit ($S$=1) instead of a qubit. In such a case,
these generalized transformations still allow one to simulate fermions
or particles with arbitrary statistics.

Two additional comments are in order: The mapping for $a_j^{\;}$ and
$a^{\dagger}_j$ described above corresponds to a one-dimensional array
of spins. The extension to higher spatial dimensions can be done
\cite{fradkin,huerta,ours} in various ways. A straightforward extension
to two dimensions is to re-map the sites of a two dimensional array
onto a one dimensional string and proceed as before. Also there is
nothing special about using the fermion instead of a quantum spin as an
alternative model of computation. One could have  just as well used the
hard core boson \cite{ours}. The main question is whether different
algebras admit a physical realization. For hard-core bosons this
realization is He$^4$ atoms.

\subsection{Evolution}

Given a fermion model algorithm, it is necessary to efficiently
obtain a corresponding standard model algorithm that at least
approximates the desired evolution. The general principle is to
map the time-dependent fermion Hamiltonian $H(t)=\sum_i H_i$ to
the standard model operators via the Jordan-Wigner transformation,
express the result in terms of a sum of simple products of Pauli
operators, and then use the Trotter approximation
\begin{equation}
 e^{-i\Delta t (H_0+H_1+\ldots)/\hbar} =
              \prod_i e^{-i\Delta t H_i/\hbar} + {\cal
              O}((\Delta t)^2).
\end{equation}
Each time step $\Delta t$ is chosen so that the final error of the
simulation is sufficiently small. Provided that the number of terms in
the sum is polynomially bounded in the number $n$ of qubits or
fermionic modes and provided that each term can be polynomially
simulated, the simulation is efficient in $n$ and $1/\mbox{error}$.

To see how to do the simulation, consider the example of the bilinear
operator $H_c = a_1^{\;} a_j^\dagger + a_j^{\;} a_1^\dagger$ in the
control Hamiltonian of the fermion model:
\begin{eqnarray*}
H_c &=& (-1)^j [\sigma^1_-\sigma_z^1 \cdots
\sigma_z^{j-1}\sigma^j_+ +
\sigma_z^1\sigma_+^1\sigma_z^2 \cdots \sigma_z^{j-1}\sigma^j_-]\\
&=&\frac{(-1)^j}{2} [ \sigma_x^1\sigma_z^2 \cdots\sigma_z^{j-1} \sigma^j_x +
\sigma_y^1\sigma_z^2 \cdots \sigma_z^{j-1}\sigma^j_y ] \ .
\end{eqnarray*}
It is readily checked that the Jordan-Wigner transformation for
the other terms in the control Hamiltonians are also decomposable
into sums of a few products of Pauli operators.

The whole idea of a quantum computer is simulating the operations
we want by using unitary matrices $U=\exp(-i\Delta t H_P/\hbar)$.
These unitary matrices, representing quantum gates, perform
reversible computation and are case-dependent. For our particular
case, we know how to simulate $H=\sigma^1_z$ in the
spin-$\frac{1}{2}$ case (it is directly implemented in the
standard model), so we ask what set of unitary operations produce
the evolution $\tilde U=\exp(-i \Delta t H_c/\hbar)$. In other
words, how do we write a $U=U_1 \dots U_k$ such that
$H_c=U^\dagger H U$? Consider for example the Hamiltonian
$H_x=\sigma_x^1\sigma_z^2 \cdots\sigma_z^{j-1} \sigma^j_x$. The
procedure is as follows: The unitary operator
\[
U_1=e^{i{\pi\over 4}\sigma_y^1} = \frac{1}{\sqrt{2}} \left
[\hat{\one} + i \sigma_y^1 \right]= \frac{1}{\sqrt{2}} \pmatrix{
1&1 \cr -1&1 \cr} \otimes \one \otimes \cdots \otimes \one
\]
takes $\sigma^1_z \rightarrow \sigma^1_x$, i.e., $U_1^\dagger
\sigma_z^1 U_1 = \sigma_x^1$. The operator
\[
U_2=e^{i{\pi\over 4}\sigma_z^1\sigma_z^2}= \frac{1}{\sqrt{2}} \left [\hat{\one} +
i \sigma_z^1 \sigma_z^2 \right]
\]
takes $\sigma^1_x \rightarrow \sigma^1_y\sigma^2_z$.
The next step is
\[
U_3=e^{i{\pi\over 4}\sigma_z^1\sigma_z^3}
\]
to take $\sigma^1_y\sigma^2_z\rightarrow
-\sigma^1_x\sigma^2_z\sigma^3_z$. By successively similar steps we
easily build the required string of operators:
$\sigma_x^1\sigma_z^2 \cdots\sigma_z^{j-1} \sigma^j_x$.

If $j$ is odd,
\[
U_j=e^{i{\pi\over 4}\sigma_z^1\sigma_z^j}
\]
will take $\sigma^1_y \sigma_z^2 \cdots \sigma_z^{j-1} \rightarrow
(-1)^{\left [ \frac{j-1}{2} \right ]} \sigma^1_x \sigma^2_z \cdots
\sigma^j_z$, where $\left [ \frac{m}{l} \right ]$ is the integer part of
$m/l$. The final operator
\[
U_{j+1}=e^{i{\pi\over 4}\sigma_y^{j}}
\]
will bring the control operator to the desired one (up to a global phase
$(-1)^{\left [ \frac{j-1}{2} \right ]}$):
\[
\sigma^1_x \sigma^2_z \cdots \sigma^{j-1}_z \sigma^j_x  \ .
\]
If $j$ is even, we need an additional unitary operator that flips the
first qubit's $\sigma_y^1$ into a $\sigma_x^1$. This flip is achieved
with the operator
\[
U_{j+2}=e^{-i{\pi\over 4}\sigma_z^{1}} = \pmatrix{ e^{-i
\frac{\pi}{4}}&0 \cr 0& e^{i \frac{\pi}{4}} \cr} \otimes \one
\otimes \cdots \otimes \one
\]
that takes $\sigma_y^1 \rightarrow \sigma_x^1$.

Hence, to construct this non-local fermion operator from the standard
model requires additional steps that are proportional to $j$ . This
number scales polynomially with the complexity so the construction is
efficient if the standard model is efficient.

The one and two-body nature of naturally occurring interactions means
that a term in a second-quantized representation of a Hamiltonian only
has one of two forms: either $a_i^\dagger a_j^{}$ or $a_i^\dagger
a_j^{}a_k^\dagger a_l^{}$. We just demonstrated how to handle the first
case. The second  case merely requires applying that algorithm
twice. This squares the complexity.






\subsection{State preparation}

In this Section we discuss the preparation of states of physical
relevance. Clearly, the preparation of the initial state is a very
important step since the study and efficiency of the given physical
process one wants to simulate depends upon it.

Consider a system of $N_e$ fermions and $n$ operators
$a^\dagger_j$ (single particle states). A generic $N_e$-particle
state of a Hilbert space ${\cal H}_{N_e}$ of antisymmetrized wave
functions can always be expanded in terms of the antisymmetric
states
\[
| \Phi_\alpha \rangle = \prod_{j=1}^{N_e} b^\dagger_j \ |{\rm vac} \rangle \ ,
\]
where $b^\dagger_j$ creates a state $j$ and $|{\rm vac} \rangle
=|0\rangle\otimes|0\rangle\cdots\otimes|0\rangle$ is the vacuum state
(i.e., $b^{\;}_j |0\rangle = 0$, $\forall j$). The operator $b^\dagger$
is in general a linear combination of $a^\dagger$'s, i.e., $b_j^\dagger
= \sum_{i=1}^n a_i^\dagger P_{ij}$ where $P_{ij}$ is some matrix and
$N_e \le n$.

The states $| \Phi_\alpha \rangle$ ($\alpha=1,\cdots,\pmatrix{n
\cr N_e \cr}$) in general form an overcomplete set of non-orthogonal states
that span the whole ${\cal H}_{N_e}$, i.e., redundantly generate
${\cal H}_{N_e}$. They are known as Slater determinants
\cite{negele:qc1988a}. Typically, $| \Phi_\alpha \rangle$ is the
result of a self-consistent mean-field (or generalized
Hartree-Fock) calculation. Even a Bardeen-Cooper-Schrieffer
superconducting state, which does not preserve the number of
particles, can be written in this way after an appropriate
canonical transformation which redefines the vacuum state
\cite{guerrero:qc1999a}.

One can easily prepare the states $| \Phi_\alpha \rangle$ noticing
that the quantum gate, represented by the unitary operator
\[
U_m = e^{i \frac{\pi}{2} (b^{\;}_m + b^\dagger_m)}
\]
when acting on the vacuum state, produces $b^\dagger_m \ | 0 \rangle$
up to a phase $e^{i \frac{\pi}{2}}$. Therefore, the successive
application of similar unitary operators will generate the state
$|\Phi_\alpha \rangle$ up to a global phase.

Except for very small systems the total Hilbert space is too large
to be fully used (it has an exponential growth with increasing
system size). In practice, one works in a subspace of ${\cal
H}_{N_e}$ that closely represents the physical state one is trying
to simulate. Generically, as initial state, we will consider a
very general expression of a many-fermion state:
\[
| \Psi (t=0) \rangle = \sum_{\alpha=1}^N {\sf a}_{\alpha} \ | \Phi_{\alpha}
\rangle \ ,
\]
where the integer $N$ is a finite and small number.  The state can be
prepared efficiently (in $N$) by a number of procedures. We now
describe one.

To make the description simple, we will assume $\sum_{\alpha=1}^N
|{\sf a}_{\alpha}|^2 =1$ and $\langle\Phi_{\alpha}
|\Phi_{\beta}\rangle = \delta_{\alpha\beta}$, which is equilvalent to
requiring $\{|\Phi_\alpha\rangle\}$ to be an orthonormal set and
$|\Psi (t=0)\rangle$ to be normalized to unity. With these assumptions the
steps of the state preparation algorithm are:
\begin{enumerate}
\item Adjoin $N$ auxiliary (ancilla) qubits, each in the state
$|0\rangle$, to the vacuum of the physical system. The resulting state is
\begin{equation}
\underbrace{|0\rangle\otimes|0\rangle\otimes\cdots|0\rangle}_N
\otimes|{\rm vac}\rangle
\equiv |0\rangle_{\sf a}\otimes|{\rm vac}\rangle
\end{equation}
\item From this state generate $\sum_{\alpha=1}^N{\sf
a}_\alpha|\alpha\rangle\otimes|{\rm vac}\rangle$ where
$|\alpha\rangle$ is an ancilla state with only the $\alpha$ qubit
being $|1\rangle$. The procedure for generating this combination of
states is described below.
\item For each $\alpha=1,\ldots,N$, conditional on the $\alpha$ qubit
being $\ket{1}$, apply the state preparation procedure for
$\ket{\Phi_{\alpha}}$.  The resulting state is
\begin{equation}
\sum_{\alpha=1}^N{\sf a}_\alpha|\alpha\rangle\otimes|\Phi_\alpha\rangle
\end{equation}
\item From this state generate
\begin{equation}\label{eq:step4}
\frac{1}{\sqrt{N}}\sum_{\alpha=1}^N{\sf a}_\alpha|0\rangle_{\sf a}
\otimes|\Phi_\alpha\rangle + {\rm terms\ without\ }|0\rangle_{\sf a}
\end{equation}
This step will also be described below.
\end{enumerate}
The final step is accepted if a measurement shows all the ancillas
being returned to $|0\rangle_{\sf a}$.  The probability of successful
preparation is thus $\sum_{\alpha=1}^N |{\sf
a}_{\alpha}|^2/N=1/N$. Consequently, on average, $N$ trials will be
needed before a successful state preparation.

The procedure to produce step 2 is most easily described by
example. We will assume $N=2$. The problem then is to generate ${\sf
a}_1|10\rangle\otimes|{\rm vac}\rangle +{\sf
a}_2|01\rangle\otimes|{\rm vac}\rangle$ from $|00\rangle\otimes|{\rm
vac}\rangle$. In what follows all operations will be only on the ancilla part
of the initial state so we will not explicitly show the vacuum. We
also note that one can always apply a rotation to a given qubit that
will take $|0\rangle$ into $x|0\rangle + y|1\rangle$ with $|x|^2
+|y|^2 =1$. The steps of the procedure are:
\begin{enumerate}
\item[2.1] Adjoin an ancilla qubit $|{\sf b}\rangle$
initially being $|0\rangle$. The initial state is now
$|0\rangle\otimes|00\rangle$.
\item[2.2] Conditional on $|{\sf b}\rangle=|0\rangle$, rotate the
$\alpha=1$ qubit, and then conditional on the $\alpha=1$ qubit being
$|1\rangle$,
flip $|{\sf b}\rangle$: 
\begin{equation}
x_1|0\rangle\otimes|00\rangle+y_1|1\rangle\otimes|10\rangle
\end{equation}
\item[2.3] Conditional on $|{\sf b}\rangle$ being $|0\rangle$, rotate the
$\alpha=2$ qubit, and then conditional on the $\alpha=2$ qubit being
$|1\rangle$, flip $|{\sf b}\rangle$:
\begin{equation}
x_1x_2|0\rangle\otimes|00\rangle+x_1y_2|1\rangle\otimes|01\rangle
+y_1|1\rangle\otimes|10\rangle
\end{equation}
\item[2.4] Project out the states with $|{\sf b}\rangle$ being $|1\rangle$:
\begin{equation}
x_1y_2|01\rangle+y_1|10\rangle
\end{equation}
\end{enumerate}
The rotations are chosen so that ${\sf a}_1=y_1$ and ${\sf a}_2=x_1y_2$.

For the explanation of step 4, we will display the physical
states. The problem is: Starting with ${\sf
a}_1|10\rangle\otimes|\Phi_1\rangle+{\sf
a}_2|01\rangle\otimes|\Phi_2\rangle$, produce (\ref{eq:step4}).
\begin{enumerate}
\item[4.1] Adjoin an ancilla qubit $|{\sf b}\rangle$
initially being $|0\rangle$. The initial state is now ${\sf
a}_1|0\rangle\otimes|10\rangle\otimes|\Phi_1\rangle + {\sf
a}_2|0\rangle\otimes|01\rangle\otimes|\Phi_2\rangle$.
\item[4.2] Conditional on $|{\sf b}\rangle$ being $|0\rangle$, rotate the
$\alpha=1$ qubit, and then conditional on the
$\alpha=1$ qubit being $|1\rangle$, flip $|{\sf b}\rangle$: 
\begin{equation}
{\sf a}_1(x_1|0\rangle\otimes|00\rangle
+y_1|1\rangle\otimes|10\rangle)\otimes|\Phi_1\rangle 
+ {\sf a}_2(x_1|0\rangle\otimes|01\rangle
+y_1|1\rangle\otimes|11\rangle)\otimes|\Phi_2\rangle
\end{equation}
\item[4.3] Conditional on $|{\sf b}\rangle$ being $|0\rangle$, rotate the
$\alpha=2$ qubit, and then conditional on the $\alpha=2$ qubit being
$|1\rangle$, flip $|{\sf b}\rangle$:
\begin{equation}
\hspace*{-1cm}
{\sf a}_1 x_1 (x_2|0\rangle\otimes|00\rangle
+y_2|1\rangle\otimes|01\rangle)\otimes|\Phi_1\rangle 
+{\sf a}_2 x_1 (x_2|0\rangle\otimes|00\rangle
+y_2|1\rangle\otimes|01\rangle)\otimes|\Phi_2\rangle
\end{equation}
\item[4.4] Project out the states with $|{\sf b}\rangle=|0\rangle$:
\begin{equation}
x_1 x_2 ({\sf a}_1|00\rangle\otimes|\Phi_1\rangle
+{\sf a}_2|00\rangle\otimes|\Phi_2\rangle)
\end{equation}
\end{enumerate}
The rotations are chosen so that $x_1 x_2$ equals $1/\sqrt{N}$
where $N=2$. Comparing step 2 with step 4, one sees they are
structurally identical, differing by the set of amplitudes generated
and the complementarity of the subspaces selected for the final
result. This latter difference in some sense makes one procedure the
inverse of the other. For the case of $N>2$, one simply replaces steps 
2.2 and 2.3 and steps 4.2 and 4.3 by ``do loops'' over $\alpha$ from 1
to $N$.

On average, the entire procedure needs $N$ trials before a successful
state preparation. (In many cases, the other measurement outcomes can
be used also to avoid too many trials.)  Construction of the initial
state thus scales as ${\cal O}(N^2 n N_e)
\le {\cal O}(N^2 n^2)$ so unless the number of Slater determinants is
exponentially large, general many-fermion states can be initialized
efficiently.



\subsection{Measurement}
\label{measurement}
 
While there is a variety of physical observables one measures
experimentally and calculates theoretically, at this time it is
difficult to demonstrate that they all can be computed efficiently on a
quantum computer. Fortunately, we will now argue that one important
class of observables, the temporal correlation functions $C_{AB}(t)$,
can be computed not only efficiently but also accurately. These
functions describe the temporal evolution of some observable $A(t)$ in
response to some weak external stimulus that couples to the system's
variable $B(0)$. They are at the heart of understanding, for example,
the optical properties of materials.

The goal is to determine correlations of the form $C_{AB}(t) =
\langle A(t) B(0) \rangle = \langle e^{i H t} A e^{-i H
t}B\rangle$ up to a sufficiently small statistical error. Clearly,
measuring efficiently $C_{AB}$ is not possible for an arbitrary
$A$ and $B$. One sufficient condition is that $A$ and $B$ are
efficiently simulatable Hamiltonians. This observation is based on
a method for determining $C_{AB}$ refined by
Kitaev~\cite{kitaev:qc1995a} and applied to the measurement of
correlation functions by Terhal and
DiVincenzo~\cite{terhal:qc1998a}. Here we give a different method
based on an idea given in~\cite{knill:qc1998c}.

A general principle that can be used to obtain $C_{AB}$ is to decompose
the operator whose expectation needs to be determined , i.e., $A(t)
B(0)$, into a small sum of operators of a simpler form and measure each
summand individually. Our method directly measures expectations of the
form $\langle U^\dagger V\rangle$ when algorithms for implementing the
unitary operators $U$ and $V$ are available. General correlation
functions are then determined by decomposing operators using a unitary
operator basis, for example the one consisting of products of Pauli
matrices.

The method for measuring $\langle U^\dagger V\rangle$ consists of
the following steps:
\begin{itemize}
\item[1.]
Adjoin via a direct product an ancilla (i.e., an auxiliary) qubit
$\bold{a}$ in the state $(\ket{0}+\ket{1})/\sqrt{2}$ with density
matrix $\rho_{\bold{a}}=(\one+\sigma^{\bold{a}}_x)/2$ to the initial
state of the system described by the density matrix $\rho$.
\item[2.]
Apply the conditional evolutions $\bar{U}_1=\ket{0}\bra{0}\tensor
U+\ket{1}\bra{1}\tensor \one$ and $\bar{U}_2=\ket{1}\bra{1}\tensor V
+\ket{0}\bra{0}\tensor \one$ ($\bar{U}=\bar{U}_1 \bar{U}_2$). The
methods of~\cite{barenco:qc1995a} may be used to implement these
evolutions given algorithms for $U$ and $V$.
\item[3.]
Measure $2 \sigma_+^{\bold{a}}=\sigma^{\bold{a}}_x+i\sigma^{\bold{a}}_y =
2\ket{0}\bra{1}$. This may be done by measuring $\sigma^{\bold{a}}_x$
and $\sigma^{\bold{a}}_y$ in sufficiently many independent trials of
these steps.
\item[4.]
Given the initial density matrix $\rho$, the expectation
\begin{eqnarray}
\langle\sigma^{\bold{a}}_x+i\sigma^{\bold{a}}_y\rangle_{\rho_{\bold{a}}
\tensor \rho}&=& 2{\rm Tr}_{n+1}[\bar{U}^\dagger
\ket{0}\bra{1}\bar{U}\rho_{\bold{a}} \tensor \rho] \\
&=& {\rm Tr}_{n+1} [\ket{0}\bra{1}\tensor U^\dagger V
\rho_{\bold{a}}\tensor\rho]={\rm Tr}_n [U^\dagger V \rho]= \langle
U^\dagger V\rangle_\rho \ ,
\end{eqnarray}
as desired. The statistical noise in the measurement of $\langle
U^\dagger V\rangle_\rho$ is determined by that of two binary random
variables and therefore depends only on the value of ${\rm Tr}_n
[U^\dagger V \rho]$, which is inside the unit complex circle. As a
result it is a simple matter to determine the number of measurement
attempts required to achieve sufficient statistical accuracy.
\end{itemize}

The procedure for measuring $C_{AB}(t)$ can now be summarized as
follows: First express $A=A(0)$ and $B=B(0)$ as a sum of unitary
operators $\displaystyle A=\sum_{j=1}^{m_A} A_j$ and $\displaystyle
B=\sum_{j'=1}^{m_B} B_{j'}$. A convenient unitary operator basis that
works well for the local observables of interest consists of all the
products of Pauli operators, as each such product is easily implemented
as a quantum algorithm.  Then, for each $j$ and $j'$ one uses the just
described method with $U= e^{iHt}A_{j}^\dagger e^{-iHt}$ and $V=B_{j'}$
to obtain $\langle A_j(t)B_{j'}(0)\rangle$. $V$ may be implemented by
simulating the evolution under $H$, applying $B_{j'}$, and then undoing
the evolution under $H$.

An alternative approach to the measurement process is von Neumann's
projection method. We sketch it here for completeness and comparison.
In this approach we also add an auxiliary (ancilla) degree of freedom to
the problem. Suppose that this extra qubit corresponds to an harmonic
oscillator degree of freedom $| e \rangle$. Then, we consider the
composite state
\[
| \Psi \rangle_S \otimes | e \rangle_0 \ ,
\]
where $| \Psi \rangle_S = \sum_j \lambda_j \ | \phi_j \rangle_S$ is the
state of the system we want to probe and $| e \rangle_t$ is the state of
the harmonic oscillator in the coordinate ($x$-)representation. The
corresponding state in the momentum ($p$-)representation is denoted $|
\hat{e} \rangle_t$.

Assume the observable ($t$-independent Hermitian operator) we want
to measure is ${\cal A}$. Then, we are interested in determining
$_S\langle \Psi | {\cal A} | \Psi \rangle_S$ in an efficient way.
Suppose that we know how to implement the unitary operation
$U^S(t) = e^{-i {\cal A} t}$. Following Kitaev we want to
implement the following conditional evolution
\[
{\cal U} = \sum_t | e \rangle_t  \ _t\langle e | \ U^S(t) \ .
\]
From the spectral theorem we can write ${\cal A} = \sum_j \Lambda_j \ |
\phi_j \rangle_S  \ _S\langle \phi_j |$. Then,
\[
{\cal U} \ | \phi_j \rangle_S \otimes | \hat{0} \rangle_0 = \sum_t {\cal
U} \ | \phi_j \rangle_S \otimes | e \rangle_t = \sum_t  e^{-i
\Lambda_j t} \ | \phi_j \rangle_S \otimes | e \rangle_t = | \phi_j
\rangle_S \otimes | \hat{\Lambda}_j \rangle_t \ ,
\]
where $| \hat{0} \rangle_0$ is a state with ($p=0$) zero momentum.
Basically, the conditional evolution ${\cal U}$ is a momentum
translation operator for the harmonic oscillator extra state.
Finally,
\[
{\cal U} | \Psi \rangle_S \otimes | \hat{0} \rangle_0 = \sum_j
\lambda_j \ |\phi_j \rangle_S \otimes | \hat{\Lambda}_j \rangle_t \ .
\]

Although the second measurement method is conceptually simpler, it
requires approximately implementing the ancillary harmonic oscillator,
the conditional evolutions for many different choices of $t$, and a
more complex analysis of the measurement statistics. The conditional
evolutions can be simplified somewhat, and in special cases (such as
as a subroutine of factoring) become very
efficient---see~\cite{kitaev:qc1995a}.


\subsection{Measurement Noise Control}


The quantum physics simulation algorithm described above is approximate
and the output is noisy. In order to properly use it, we need to have
explicit estimates of the error $\epsilon$ in the inferred expectations
given the noise in the implementation. Furthermore, the effort required
to make $\epsilon$ small must scale polynomially with $1/\epsilon$.
There are three sources of error that need to be considered. The first
is associated with intrinsic noise in the implementation of the gates
due to imperfections and unwanted interactions. The second comes from
the discretization of the evolution operator and the use of the Trotter
decomposition. The third is due to the statistics in measuring the
desired correlation function using the technique given above.

\subsubsection{Gate imperfections}

The problem of gate imperfections can be dealt with by using quantum
error correction~\cite{shor:qc1995b,steane:qc1995a} and fault tolerant
quantum computation
\cite{shor:qc1996a,aharonov:qc1996a,kitaev:qc1996a,knill:qc1998a,preskill:qc1998a}.
According to the accuracy threshold theorem, provided the physical
gates have sufficiently low error, it is possible to quantum compute
arbitrarily accurately. The fault tolerant computation implements
unitary operations and measurements on encoded qubits with overheads
bounded by ${\cal O}(\log^k(1/\epsilon))$ for some $k$. This exponentially
efficient convergence implies that the effects of physical noise can in
principle be ignored.

\subsubsection{Discretization error}

A second type of error is the one introduced by the discretization of
the evolution operator. This discretization is very similar to the one
used in classical simulation of dynamical quantum systems. It is
possible to estimate the size of this error by a detailed analysis of
the discretization. For example using the Trotter approximation
\[
e^{-i(H_1+H_2)\Delta t}=e^{-i(H_1)\Delta t/2}
e^{-i(H_2)\Delta t}e^{-i(H_1)\Delta t/2} + {\cal O}((\Delta t)^3).
\]
The  coefficient of $(\Delta t)^3\sim -i(H_1+H_2)^3/6 $ can be bounded by
estimating the largest eigenvalue of $H_1$ and $H_2$.


\subsubsection{Measurement statistics}

Our technique for measuring the correlation function $\langle A(t) B(0)
\rangle$ requires measuring the expectations of unitary operators
$U_j^\dagger V^{\;}_{j'}$ associated with the implemented evolution. In
most cases, the operators $A$ and $B$ are a sum of ${\cal O}(m_{A,B})$
products of Pauli matrices, so that ${\cal O}(m_A)$ $U^\dagger_j$'s and ${\cal
O}(m_B)$ $V_{j'}$'s are needed. This means that the expectation is a sum
of ${\cal O}(m_A m_B)$ random variables $r_{jj'}$, where $|r_{jj'}|\leq
1$. To assure that the statistical noise (given by the standard
deviation) is less than $\epsilon$ it suffices to measure each
$r_{jj'}$ ${\cal O}(m_A m_B/\epsilon^2)$ times.

\section{Example: Resonant Impurity Scattering}
\label{section4}

\subsection{Formulation of the Physical Problem}

Our toy problem is a ring of $n$ equally-spaced
lattice sites on which spinless fermions hop to nearest neighbor
sites or hop onto or from an ``impurity'' state. The length of the ring
is $L = n a$, where $a$ is the distance between sites. The system is
described by the Hamiltonian (in second quantized form)
\begin{equation}
H = - {\cal T} \sum_{i=1}^{n} ( c_i^{\dagger} c_{i+1}^{\;} +
c_{i+1}^{\dagger} c_i^{\;} ) + \epsilon \ b^{\dagger} b +
\frac{V}{\sqrt{n}} \sum_{i=1}^{n} (c_i^{\dagger} b + b^{\dagger}
c_i^{\;}) \ ,
\label{toy}
\end{equation}
As usual, $b$'s and $c$'s are fermion (anticommuting) operators. The
index $i$ labels the lattice sites ($R_i = i a$ is the lattice site
position) and strict periodic boundary conditions are assumed, i.e.,
\begin{equation}
c_{i+n}^{\dagger} = c_i^{\dagger} \ .
\end{equation}

We now imagine that the system is initially prepared in the zero
temperature ground state of the ring in the
absence of the impurity. Then, at time $t=0$, a fermion is injected
into the impurity state. After the system has evolved
for some time $t$, we want to compute the probability amplitude that
the evolved state is still in the initial state. The relevant quantity
to compute is ($\hbar =1$ and $t \geq 0$)
\begin{eqnarray}
G(t) &=& \langle \Psi(0) | b(t) b^{\dagger}(0) | \Psi(0) \rangle \ , \\
b(t) &=& e^{i H t} \ b(0) \ e^{-i H t} \ ,
\end{eqnarray}
where the initial state is the Fermi sea of $N_e \leq n$ fermions
\begin{equation}
| \Psi(0) \rangle = | FS \rangle = \prod_{i=0}^{N_e-1} c^{\dagger}_{k_i}
| 0 \rangle \ .
\end{equation}
$| 0 \rangle$ is the vacuum of fermions and
\begin{equation}
c^{\dagger}_{k_i} = \frac{1}{\sqrt{n}} \sum_{j=1}^n \ e^{i k_i R_j} \
c^{\dagger}_{j} \ .
\end{equation}
The wave number $k_j$ is determined from the periodic boundary
conditions, $c_{i+n}^{\dagger} = c_i^{\dagger}$, which implies
\begin{equation}
k_j = \frac{2 \pi n_j}{L} \ , \mbox{ with $n_j$ an integer} \ .
\end{equation}
There is no unique way to choose the set of $n_j$'s. The common
convention is to define the first Brillouin zone as
\begin{equation}
-\frac{\pi}{a} < k \leq \frac{\pi}{a} \ ,
\end{equation}
with $k$ values uniformly distributed in this interval with spacing
$\Delta k = 2 \pi/L$.

\subsection{Quantum Algorithm}

We want to write a quantum algorithm that allows one to compute $G(t)$
(see the Appendix for the exact closed-form solution). To this end, we
start by representing fermion operators in Eq. \ref{wave} in terms of
Pauli matrices. Because of the form of the hybridization term, Eq.
\ref{hybrid}, a most convenient representation is the following
\begin{eqnarray}
b = \sigma_-^1 & b^\dagger = \sigma_+^1  \\
c^{\;}_{k_0} = - \sigma_z^1 \sigma_-^2  & c^{\dagger}_{k_0} = -
\sigma_z^1 \sigma_+^2 \\
 \vdots  &  \vdots \\
c^{\;}_{k_{n-1}} = (-1)^{n} \sigma_z^1 \sigma_z^2 \cdots \sigma_z^{n}
\sigma_-^{n+1}  & \qquad c^{\dagger}_{k_{n-1}} = (-1)^{n} \sigma_z^1
\sigma_z^2 \cdots \sigma_z^{n} \sigma_+^{n+1} ,\\
\end{eqnarray}
from which the following mapping results
\begin{eqnarray}
b^\dagger b^{\;} &=& 2 (\one + \sigma_z^1)\nonumber \\
c^{\dagger}_{k_i} c^{\;}_{k_i}&=& 2 (\one + \sigma_z^{i+2})\nonumber \\
c^{\dagger}_{k_0} b^{\;} + b^\dagger c^{\;}_{k_0}&=& \frac{1}{2}
(\sigma_x^1 \sigma_x^2 + \sigma_y^1 \sigma_y^2) \nonumber \ .
\end{eqnarray}
Therefore, the Hamiltonian operator reads
\begin{equation}
\frac{H}{2} =\left [ \epsilon + \sum_{i=0}^{n-1} {\cal E}_{k_i} \right ] \one +
\epsilon \sigma_z^1 + \sum_{i=0}^{n-1} {\cal E}_{k_i}\sigma_z^{i+2}
+ \frac{V}{4} (\sigma_x^1 \sigma_x^2 + \sigma_y^1 \sigma_y^2) \ .
\end{equation}
An additional simplification can be introduced when one realizes that the
structure of the observable to be measured is such that
\begin{equation}
b(t) = e^{i H t} b(0) e^{-i H t} = e^{i \bar{H} t} \sigma_-^1 e^{-i
\bar{H} t} \ ,
\end{equation}
where $\bar{H}$ is given by
\begin{equation}
\bar{H} = 2 \epsilon \sigma_z^1 + 2 {\cal E}_{k_0} \sigma_z^2 +
\frac{V}{2} (\sigma_x^1 \sigma_x^2 + \sigma_y^1 \sigma_y^2) \ ,
\end{equation}
and, therefore, the ``string'' one has to simulate has length equal to
two (it involves only qubits 1 and 2)
\begin{equation}
{\cal A}(t) = b(t)b^\dagger(0) = e^{i \bar{H} t} \sigma_-^1 e^{-i
\bar{H} t} \sigma_+^1 \ .
\end{equation}
If we were to transform $\bar{H}= U H_{P1} U^\dagger$ unitarily with $U
= \prod_{j=1}^n e^{i H_{P2}^j t_j}$ and $n$ a finite integer ($U
U^\dagger = \one$) in such a way that both $H_{P1}$ and $H_{P2}$ are
physical Hamiltonians, then the simulation would be straightforward.
(We call this type of mapping a physical unitary mapping.) For our two
qubit case, one can always perform a physical unitary mapping with
\begin{equation}
U = e^{i \frac{\pi}{4} \sigma^2_x}  e^{-i \frac{\pi}{4} \sigma^1_y}
e^{-i \frac{\theta}{2} \sigma^1_z \sigma^2_z} e^{i
\frac{\pi}{4}\sigma^1_y} e^{i \frac{\pi}{4} \sigma^1_x}  e^{-i
\frac{\pi}{4}\sigma^2_x} e^{-i \frac{\pi}{4} \sigma^2_y}  e^{i
\frac{\theta}{2}\sigma^1_z \sigma^2_z} e^{-i \frac{\pi}{4} \sigma^1_x}
e^{i \frac{\pi}{4} \sigma^2_y} \ ,
\end{equation}
\begin{equation}
H_{P1} = \frac{1}{2} (E - \sqrt{\Delta^2+V^2}) \sigma^1_z +
\frac{1}{2} (E + \sqrt{\Delta^2+V^2}) \sigma^2_z \ ,
\end{equation}
with $E=2(\epsilon + {\cal E}_{k_0})$, $\Delta=2(\epsilon - {\cal
E}_{k_0})$, and $\cos \theta = 1/\sqrt{1+\delta^2}$ with
$\delta=(\Delta+\sqrt{\Delta^2+V^2})/V$.

In general, such a constrained transformation is not easily realized
and one performs a Trotter decomposition
\begin{equation}
e^{i \bar{H} t} = \left [ e^{i \bar{H} s} \right ]^M = \left [ e^{i
\bar{H}_z s} e^{i \bar{H}_{xy} s} + {\cal O}(s^2) \right ]^M
\end{equation}
where $\bar{H} = \bar{H}_z + \bar{H}_{xy}$ with $\bar{H}_{xy} =
\frac{V}{2} (\sigma_x^1 \sigma_x^2 + \sigma_y^1 \sigma_y^2)$ and time
slice $s=\frac{t}{M}$. On the other hand, one can easily perform a physical
unitary mapping for
 $e^{i \bar{H}_{xy} s}$
\begin{equation}
e^{i \bar{H}_{xy} s} = \bar{U} e^{i H_{P1} s} \bar{U}^\dagger \ ,
\end{equation}
where $H_{P1} = \frac{V}{2} (\sigma_x^1 - \sigma_y^2)$ and
\begin{equation}
\bar{U} = e^{i \frac{\pi}{4} \sigma_x^2} e^{-i \frac{\pi}{4} \sigma_y^1}
e^{-i \frac{\pi}{4} \sigma_z^1 \sigma_z^2} \ .
\end{equation}
Finally, the ``string'' one has to simulate with the quantum computer is
\begin{eqnarray}
{\cal A}(t) &\simeq& \left [ S(s) \right ]^M \sigma_-^1 \left [
S^\dagger(s) \right ]^M \sigma_+^1 \nonumber \\
S(s) &=&  e^{i \bar{H}_z s} \bar{U} e^{i H_{P1} s} \bar{U}^\dagger  \ .
\end{eqnarray}
and $G(t) = \langle {\cal A}(t) \rangle$.



\section{Concluding Remarks}
\label{section5}

We investigated the implementation of algorithms for the simulation of
fermionic quantum systems, and gave an explicit mapping that relates
the usual qubit of a quantum computer to the fermionic modes that we
want to simulate. Our attention focused on the so-called sign problem.
It is a problem appearing in attempts to simulate classically the
dynamics of quantum systems. We reviewed the origin of this problem
and showed how this problem is avoided in quantum computing
simulation. The evolution of quantum computers are intrinsically
quantum mechanical and this is the main difference with a classical
computer that allows one to solve the sign problem. We studied sources
of errors in a quantum computer, such as gate imperfections and the
expansion of the evolution operator, and argued that they would not
open a back door to a problem similar to the sign problem.

We gave a very general definition of what a model of quantum
computation is. In particular and because of our particular interest,
i.e., the simulation of fermion systems, we described the standard and
the fermionic models (``Grassmann Chip''). These are, of course, not
the only ones. Isomorphisms of $*$-algebras allow one to introduce more
``esoteric'' models \cite{ours}. Indeed, there is nothing special about
the spinless fermionic model of quantum computation. One could have
used a ``hard-core boson'' model which admits, in principle, a
realization in terms of He$^4$ atoms. The key point is the
implementation of the physical gates.

Our effort focused on the simulation of the dynamics of fermionic
quantum systems. However other problems can be of interest: the
thermodynamic or ground state properties of a Hamiltonian. Even if one
had a quantum computer, it is not clear how to use it to efficiently
compute these quantities. On the other hand, at present, no proof
exists showing that this is not possible.

An approach that in principle could be used to compute the spectrum of
a Hamiltonian $H$ (e.g., the ground state) or expectation values of
arbitrary observables is the adiabatic ``switching on'' in conjunction
with the Gell-Mann-Low theorem  \cite{gellmann} of quantum field
theory. The idea simply consists of introducing a fictitious
Hamiltonian
\begin{equation}
H_\epsilon(t) = H_0 + f_\epsilon(t) \  H_1 \ ,
\end{equation}
where both $H_0$ and $H_1$ are time independent operators ($H = H_0 +
H_1$) and the scalar function $f_\epsilon(t)$ is such that $\lim_{t
\rightarrow \pm \infty} f_\epsilon(t)= 0$ and $\lim_{t \rightarrow 0}
f_\epsilon(t)= 1$, for an arbitrary adiabatic parameter $\epsilon$. In
other words, $H_\epsilon(t=0)=H$ and $H_\epsilon(t=\pm \infty)=H_0$.
$H_0$ is typically an operator whose spectrum is known, e.g., an
arbitrary bilinear operator representing a mean-field solution of $H$
and whose eigenstates can be straightforwardly prepared (let's call it
$|\Phi_0 \rangle$). The Gell-Mann-Low theorem asserts that
\begin{equation}
\lim_{\epsilon \rightarrow 0}  \frac{U_\epsilon(0,-\infty) |\Phi_0
\rangle}{\langle \Phi_0 | U_\epsilon(0,-\infty) |\Phi_0 \rangle} =
\frac{|\Psi_0 \rangle}{\langle \Phi_0|\Psi_0 \rangle}
\end{equation}
if the state whose limit one is performing admits a series expansion in
a coupling parameter characterizing the strength of $H_1$. This formal
device generates the eigenstate adiabatically connected to $|\Phi_0
\rangle$. The theorem does not guarantee that if $|\Phi_0 \rangle$ is
the ground state of $H_0$ then $|\Psi_0 \rangle$ is the ground state of
$H$. If the conditions of the theorem are satisfied then computation of
the spectrum of $H$ is straightforward. To our knowledge this approach
has never been implemented in practice.

The work presented here is only a first step in a program
investigating the simulation of quantum systems using quantum
computers. We have given a rather explicit algorithm for a simple
problem and we will increase the complexity of the problems in the work
to come. An interesting problem would be to provide algorithms to test
for superconductivity in systems such as the Hubbard model. Such
simulations using classical computers cannot unequivocally answer this
important question because of the sign problem, but a quantum computer
could.

\acknowledgements

We acknowledge the Aspen Center for Physics for its kind hospitality
while part of this work was performed.  Work at Los Alamos is sponsored
by the US DOE under contract W-7405-ENG-36.

\newpage
\appendix
\section*{Toy Problem: Exact Solution}

We can rewrite our original Hamiltonian, Eq. \ref{toy}, in the wave number
representation:

\begin{enumerate}

\item
\underline{Kinetic energy:}
\begin{equation} \hspace*{-2cm}
T = - {\cal T} \sum_{i=1}^{n} ( c_i^{\dagger} c_{i+1}^{\;} +
c_{i+1}^{\dagger} c_i^{\;} ) = -\frac{\cal T}{n} \sum_{k,k',i} e^{i (k'
- k) R_i} \left(  e^{i k' a} + e^{-i k a} \right) \ c^{\dagger}_k
c^{\;}_{k'} \ ,
\end{equation}
Thus,
\begin{equation}
T = \sum_{i=1}^n {\cal E}_{k_i} \ c^{\dagger}_{k_i}  c^{\;}_{k_i} \ ,
\end{equation}
where ${\cal E}_k = - 2 {\cal T} \cos{k a}$ \ .

\item
\underline{Hybridization energy:}
\begin{equation}
H_{hyb} =  \frac{V}{\sqrt{n}} \sum_{i=1}^{n} (c_i^{\dagger} b +
b^{\dagger} c_i^{\;}) = \frac{V}{n} \sum_{k,i} \left ( e^{-i k R_i}
\ c^{\dagger}_k b + e^{i k R_i} \ b^{\dagger} c^{\;}_k \right ) \ .
\end{equation}
Therefore,
\begin{equation}
H_{hyb} =  V \sum_{k} \left ( c^{\dagger}_k b + b^{\dagger} c^{\;}_k
\right ) \delta_{k 0} \ .
\label{hybrid}
\end{equation}
\end{enumerate}
Thus, in the wave number representation, the total Hamiltonian reads
\begin{equation}
H =  \sum_{i=0}^{n-1} {\cal E}_{k_i}^{} \ c^{\dagger}_{k_i}  c^{\;}_{k_i} +
\epsilon \ b^{\dagger} b^{\;} +  V \sum_{i=0}^{n-1} \left (
c^{\dagger}_{k_i} b + b^{\dagger} c^{\;}_{k_i} \right )
\delta_{k_i 0} \
\label{wave}
\end{equation}

\begin{figure}[htb]
\setlength{\unitlength}{1cm}
\centerline{\epsfbox{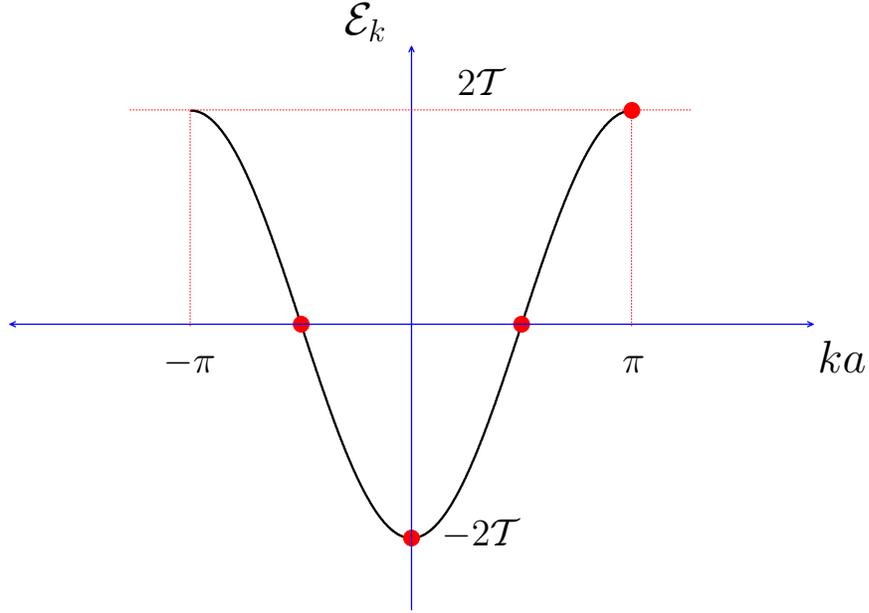}
        \put(-7.3,8){\Large ${\cal E}_k$}
        \put(-9.7,3.5){\large $-\pi$}
        \put(-3.6,3.5){\large $\pi$}
        \put(-5.8,7.2){\large $2 {\cal T}$}
        \put(-6.,1.2){\large $-2 {\cal T}$}
        \put(-1,3.5){\Large $k a$}
}
\caption{Single-particle energy spectrum for an $n=4$ site ring.}
\label{fig1}
\end{figure}

The quantization of $k$ quantizes the single particle energy spectrum
${\cal E}_k$. For example, for an $n=4$ site ring, the allowed $k$
values and energies ${\cal E}_k$ are shown in Fig. \ref{fig1}.
Its ground state with $n_e=3$ is
\begin{equation}
| FS \rangle = c^{\dagger}_{k_0} c^{\dagger}_{k_1} c^{\dagger}_{k_2} | 0
\rangle \ , \mbox{ with}\  k_0 = 0 \ , k_1 = \frac{\pi}{2} \ , \mbox{
and}\ k_2 = -\frac{\pi}{2} \ .
\end{equation}

The Heisenberg equations of motion are:
\begin{eqnarray}
\dot{b}^{\dagger} &=& i \left [ H, b^{\dagger} \right ] \\
\dot{c}^{\dagger}_k &=& i \left [ H, c^{\dagger}_k \right ] \ ,
\end{eqnarray}
where $[A,B]=AB-BA$. These commutators equal
\begin{eqnarray}
\left [ H, b^{\dagger} \right ] &=& \epsilon \ b^{\dagger} + V
c^{\dagger}_{k} \ \delta_{k 0} \\
\left [ H, c^{\dagger}_k \right ] &=& {\cal E}_k c^{\dagger}_k + V
b^{\dagger} \ \delta_{k 0} \ .
\end{eqnarray}
Clearly, one can distinguish two cases:

1) $k \neq 0$ :

\begin{equation}
c^{\dagger}_k (t) = e^{i {\cal E}_k t} \  c^{\dagger}_k (0) = e^{-i 2 t
{\cal T} \cos{k a}} \  c^{\dagger}_k (0)
\end{equation}

2) $k = 0$ (${\cal E}_0 = - 2 {\cal T}$):

\begin{eqnarray}
-i \ \dot{b}^{\dagger} &=&   \epsilon \ b^{\dagger} + V
c^{\dagger}_{0} \\
-i \ \dot{c}^{\dagger}_0 &=&  {\cal E}_0 c^{\dagger}_0 + V
b^{\dagger} \ ,
\end{eqnarray}
or in matrix notation
\begin{equation}
\pmatrix{ \dot{b}^{\dagger} \cr \dot{c}^{\dagger}_0 \cr} =
i \ \pmatrix{ \epsilon & V \cr V & {\cal E}_0 \cr} \ \cdot \
\pmatrix{ b^{\dagger} \cr  c^{\dagger}_0 \cr} \ ,
\end{equation}
The solution of these first-order differential equations are
\begin{equation}
\pmatrix{ b^{\dagger}(t) \cr c^{\dagger}_0(t) \cr} =
\exp[{i \ t \ \pmatrix{ \epsilon & V \cr V & {\cal E}_0 \cr}}] \
\cdot \ \pmatrix{ b^{\dagger}(0) \cr  c^{\dagger}_0(0)
\cr} \ .
\end{equation}
From elementary matrix algebra
\begin{equation} {\cal U}(t) =
\exp[{i \ t \ \pmatrix{ \epsilon & V \cr V & {\cal E}_0
\cr}}] = \pmatrix{ x & -y \cr y & x \cr} \
\pmatrix{ e^{i t \lambda_1} & 0 \cr 0 & e^{i t \lambda_2}
\cr} \  \pmatrix{ x & y \cr -y & x \cr}  \ ,
\end{equation}
where $\lambda_{1,2}$ are the eigenvalues of $\pmatrix{ \epsilon
& V \cr V & {\cal E}_0 \cr}$ with corresponding eigenvectors
$\pmatrix{ x \cr y \cr}$ and $ \pmatrix{ -y \cr x
\cr}$, and $x^2+y^2=1$. After elementary algebraic
manipulations:
\begin{equation}
\lambda_{1,2} = \frac{\epsilon + {\cal E}_0}{2} \pm \sqrt{ \left(
\frac{\epsilon - {\cal E}_0}{2} \right)^2 + V^2} \ ,
\end{equation}
and
\begin{eqnarray}
x &=& \frac{V}{\sqrt{(\lambda_1 - \epsilon)^2 + V^2}} \\
y &=& \frac{V}{\sqrt{(\lambda_2 - \epsilon)^2 + V^2}} \ .
\end{eqnarray}
Thus, the operator $b(t)$ is expressible as
\begin{equation}
b(t) = {\cal U}^*_{11}(t) \ b(0) + {\cal U}^*_{12}(t) \ c_0(0) \ ,
\end{equation}
where
\begin{eqnarray}
{\cal U}^*_{11}(t) &=&  x^2 e^{-i t \lambda_1} + y^2 e^{-i t \lambda_2} \\
{\cal U}^*_{12}(t) &=& x y \left( e^{-i t \lambda_1} - e^{-i t
\lambda_2} \right ) \ ,
\end{eqnarray}
and $\det{{\cal U}(t)} = e^{i t (\lambda_1+\lambda_2)} = e^{i t
(\epsilon + {\cal E}_0)}$.

\begin{figure}[htb]
\setlength{\unitlength}{1cm}
\centerline{\epsfbox{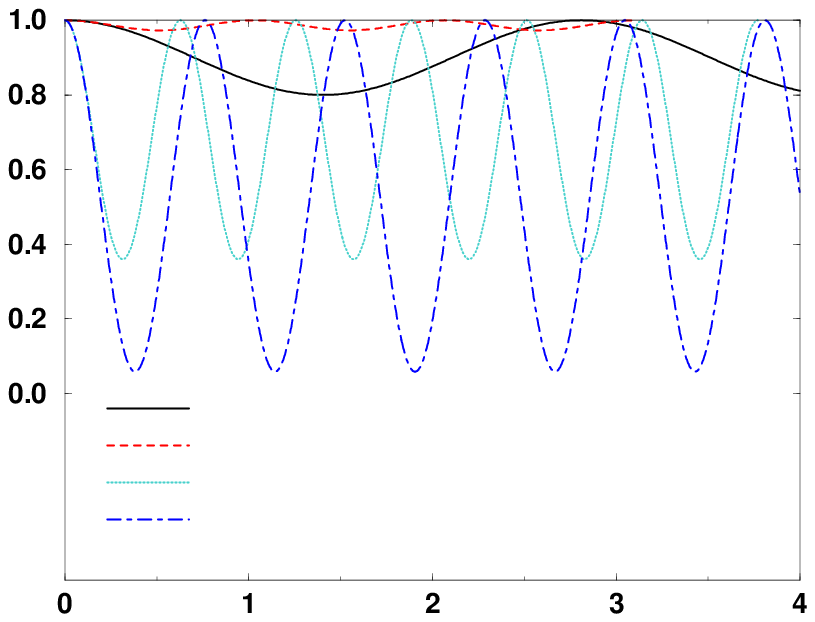}
        \put(-13,6){\Large $|G(t)|^2$}
        \put(-8.5,3.0){\large $\epsilon=0$, $V=1/2$}
        \put(-8.5,2.6){\large $\epsilon=-8$, $V=1/2$}
        \put(-8.5,2.2){\large $\epsilon=-8$, $V=4$}
        \put(-8.5,1.8){\large $\epsilon=0$, $V=4$}
       \put(-6.5,0.5){\Large $t$}
}
\caption{$|G(t)|^2$ for different values of the parameters (${\cal E}_0
= -2$).}
\label{fig2}
\end{figure}

Finally, $G(t)$ is simply given by
\begin{equation}
G(t) = {\cal U}^*_{11}(t) \ .
\end{equation}
$|G(t)|^2 = x^4+y^4+2 x^2 y^2 \cos{(\lambda_1 - \lambda_2) t}$ and is
plotted in Fig.~\ref{fig2}.
Note that $G(t)$ is independent of the number of fermions
present in the initial state $| \Psi(0) \rangle$. This is
why this toy problem reduces to a two qubit problem.



%
%
\end{document}